\documentclass[12pt]{article}

\catcode`\@=11

\global\arraycolsep=2pt
\usepackage{amsbsy,amssymb,latexsym,amsfonts,amsmath}
\usepackage{graphicx,color}

\allowdisplaybreaks

\begin{document}
\begin{flushright}
\parbox{4.2cm}
{RUP-22-15}
\end{flushright}

\vspace*{0.7cm}

\begin{center}
{ \Large If you want to cross singularity, wrap it!}
\vspace*{1.5cm}\\
{Yu Nakayama}
\end{center}
\vspace*{1.0cm}
\begin{center}

Department of Physics, Rikkyo University, Toshima, Tokyo 171-8501, Japan

\vspace{3.8cm}
\end{center}

\begin{abstract}
In two-dimensional string theory, a probe D0-brane does not see the black hole singularity due to a cancellation between its metric coupling and the dilaton coupling. A similar mechanism may work in the Schwarzschild black hole in large $D$ dimensions by considering a  suitable wrapped membrane. From the asymptotic observer, the wrapped membrane looks disappearing into nothing while the continuation of the time-like trajectory beyond the singularity suggests that it would reappear as an instantaneous space-like string stretching from the singularity. A null trajectory can be extended to a null trajectory beyond the singularity. Not only the effective particle but an effective string from the wrapped membrane can exhibit the same feature.

\end{abstract}

\thispagestyle{empty} 

\setcounter{page}{0}

\newpage

\section{Introduction}
In general relativity, different observers can observe different physics. The notion of the horizon is one particular example, where the distant observer cannot extend their time evolution, but the local observer can extend their time evolution so that he or she can cross it (without feeling any violence).  When the time evolution cannot be extended by any observer, it is called singularity. 

In general relativity, the singularity corresponds to geodesic incompleteness because the time evolution of a local observer (without any other force) is characterized by the geodesic motion. In a unified theory of gravity such as in string theory, however, the notion of the geodesic incompleteness (of one metric) may not necessarily represent the singularity because different observers can feel a different metric. For instance, the open string metric can be different from the closed string metric in string theory. D-brane can feel different metrics because of the dilaton coupling.

The two-dimensional string theory is an example of solvable quantum gravity both from the viewpoint of the worldsheet and the target space (see e.g. \cite{Nakayama:2004vk} and reference therein). The D-brane dynamics in two-dimensional string theory is of particular importance because it may give a non-perturbative formulation of the two-dimensional string theory as a dual holographic description \cite{McGreevy:2003kb}\cite{Takayanagi:2003sm}\cite{Douglas:2003up}. 

Two-dimensional string theory admits a non-trivial dilatonic black hole solution. It has an exact worldsheet conformal field theory description based on the coset model \cite{Witten:1991yr}\cite{Dijkgraaf:1991ba}. The D-brane dynamics on the two-dimensional black hole has been studied in the literature as a boundary state in the coset model \cite{Yogendran:2004dm}\cite{Nakayama:2005pk}\cite{Nakayama:2006qm}\cite{Nakayama:2007sb}. The salient features of the D-brane dynamics in the two-dimensional black hole background are tachyon-radion correspondence and the invisibility of the black hole singularity, which we will review in the main text. The invisibility of the black hole singularity of the probe D-brane is a manifestation of the observer-dependence in the notion of singularity in string theory.

More recently, the two-dimensional black hole solution in string theory has been revisited from the view of the large $D$ expansion in general relativity. In \cite{Soda:1993xc}\cite{Emparan:2013xia}, it was observed that upon dimensional reduction, the large $D$ dimensional Einstein gravity is effectively described by the two-dimensional dilaton gravity that appears in the two-dimensional string theory. In particular, in the large $D$ limit, the Schwarzschild  black hole (or Schwarzschild-Tangherlini black hole) is described by the two-dimensional black hole solution that we have alluded to. We expect that the large $D$ expansion of the general relativity and its connection to the two-dimensional string black hole may shed new light on our understanding of the Schwarzschild  black hole in quantum gravity \cite{Chen:2021emg}\cite{Balthazar:2022szl}.  See e.g. \cite{Emparan:2020inr} and reference therein for the aspects of general relativity in the large $D$ dimensions. See \cite{Chen:2021qrz}\cite{Brustein:2021cza}\cite{Giveon:2021gsc}\cite{Chen:2021dsw}\cite{Giribet:2021cpa}\cite{Giataganas:2021jbj}\cite{Brustein:2021qkj} for recent studies on this subject.

The aim of this paper is to study the large $D$ dimension analouge of the D-brane dynamics in the two-dimensional black hole in string theory. We will show that a wrapped membrane in the large $D$ Schwarzschild  black hole gives the effective dynamics that resembles the D-brane in the two-dimensional black hole. As in the two-dimensional black hole background, the trajectory of the wrapped probe membrane can be extended beyond the singularity. We will propose the fate of the wrapped membrane based on the continuation of the trajectory. We also show that a certain effective string from the wrapped membrane does not see singularity.

\section{Large $D$ black hole and $D=2$ black hole}

Consider Einstein's theory of gravity without cosmological constant in $D$ space-time dimensions. We are interested in black hole solutions with $SO(D-2)$ rotational symmetry. We postulate the  ansatz for the metric:
\begin{align}
ds^2_D = g_{\mu\nu}dx^\mu dx^\nu + r_0^2 e^{-\frac{4\Phi}{D-2}} d\Omega_{D-2}^2 \ ,
\end{align}
where $\mu=t,\rho$ and $d\Omega_{D-2}^2$ is the maximally symmetric metric for the $D-2$ dimensional sphere. We assume $g_{\mu\nu}$ and $\Phi$ only depend on two-dimensional coordinates $t$ and $\rho$.

In the large $D$ limit, the Kaluza-Klein reduction on $D-2$ dimensional sphere gives the effective two-dimensional dilaton-gravity  \cite{Soda:1993xc}\cite{Emparan:2013xia}:
\begin{align}
I = \int d^2 x \sqrt{-g} e^{-2\Phi} (R+ 4 \partial_\mu \Phi \partial^\mu \Phi + 4 k^{-1}) \ ,  
\end{align}
where $k \sim \left(\frac{2 r_0}{D}\right)^2$. Remarkably, it coincides with the effective action of the two-dimensional string theory. In particular, it has a solution of the two-dimensional black hole that admits the exact conformal field theory description as $SL(2,\mathbb{R})_k/U(1)$ coset model (in the large $k$ limit).

A relevant solution of the equations of motion describing the two-dimensional black hole is
\begin{align}
ds^2 &= k ( -\tanh^2 \rho dt^2 + d\rho^2)  \cr    
e^{-2\Phi} & = \cosh^2 \rho \ , \label{twoBH}
\end{align}
which can be uplifted to the $D$-dimensional Schwarzschild  black hole (or Schwarzschild-Tangherlini black hole) 
\begin{align}
ds^2_D = k ( -\tanh^2 \rho dt^2 + d\rho^2) + r_0^2 (\cosh^2 \rho)^{\frac{2}{D-2}} d\Omega_{D-2}^2 
\end{align}
in the large $D$ limit. The existence of the uplift means that the Kaluza-Klein reduction used here is a consistent truncation (at least for this solution). We also note that the dilaton $\Phi$ is related to the size of the sphere, justifying the relation between the entropy of the two-dimensional black hole and the value of the dilaton at the horizon.

\begin{figure}[hbtp]
 \centering
 \includegraphics[bb=0 0 200 230,keepaspectratio, scale=0.7]{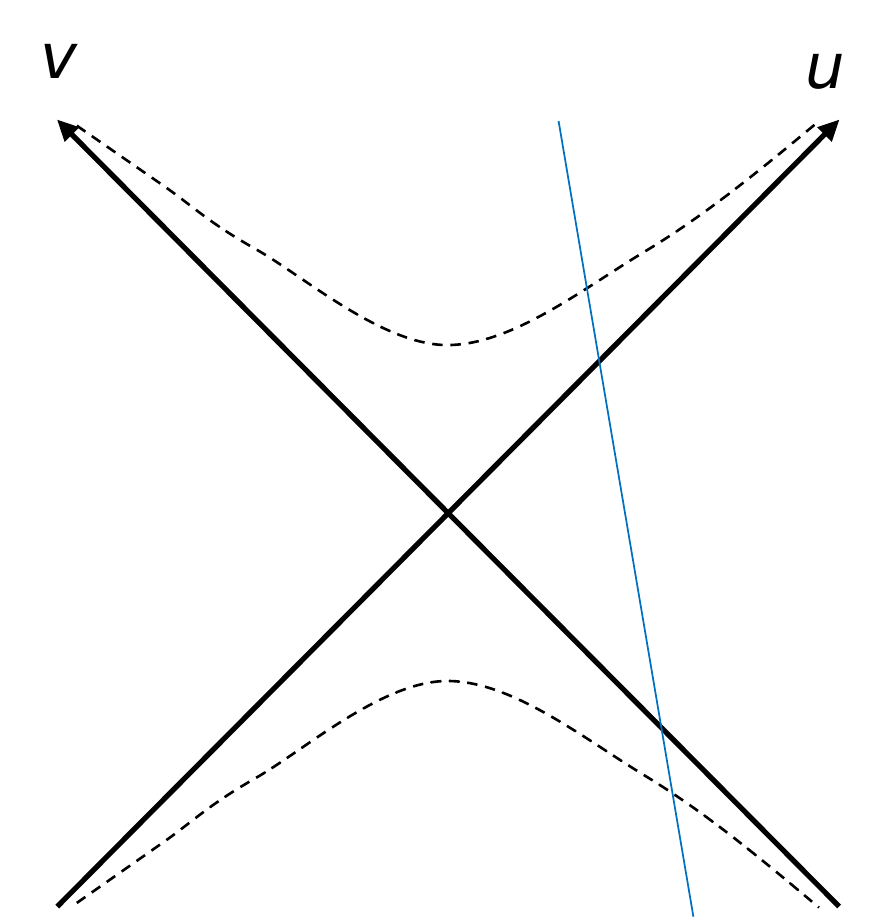}
 \caption{Kruskal diagram for the two-dimensional black hole. The singularity is located at $uv=1$ represented by dashed curves. The blue straight line is a trajectory of a D0-brane.}
 \label{fig1}
\end{figure}

Solution \eqref{twoBH} only describes the black hole outside of the event horizon located at $\rho=0$.
To extend the coordinate inside the black hole, we introduce an analogue of the Kruskal coordinate
\begin{align}
u &= \sinh \rho e^{t} \cr
v &= -\sinh \rho e^{-t} \ . 
\end{align}
so that
\begin{align}
ds^2 &= -2 k \frac{du dv}{1-uv} \cr
e^{-2\Phi} & = (1-uv) \ . \label{Kruskal}
\end{align}
The original $(t,\rho)$ coordinate covers $u>0, v<0$.
Here $uv=0$ corresponds to the event horizon and $uv=1$ corresponds to the singularity. As in the maximally extended Schwarzschild solution, it also describes the white hole region ($u<0, v<0$) and the causally disconnected asymptotic region ($u<0, v>0$). See Fig \ref{fig1}.

Let us briefly review the worldsheet description of the two-dimensional black hole. It is based on the $SL(2,\mathbb{R})_k/U(1)$ coset model. At level $k$, the $SL(2,\mathbb{R})_k/U(1)$ coset model has the central charge $c= 2 + \frac{6}{(k-2)}$. If we are interested in the two-dimensional bosonic string theory (so that the total space-time dimension is two), we may want to put $k=\frac{9}{4}$ to obtain $c=26$, which is in the strongly coupled regime $k<3$. 
Alternative use of the two-dimensional black hole solution in string theory is to consider thermal NS5-branes. The radial part of the thermal NS5-brane is effectively described by the two-dimensional black hole background \cite{Callan:1991dj}. Here $k$ is related to the number of NS5-branes. As a $SL(2,\mathbb{R})_k/U(1)$ coset model, $k$ is not quantized, but the number of flux in the sphere part makes $k$ quantized in the thermal NS5-branes.
In the correspondence with the large $D$ dimensional Schwarzschild  black hole, $k -2 = \left(\frac{2r_0}{D}\right)^2$ as observed in \cite{Chen:2021emg}.

The vertex operator describing the mode outside of the horizon can be obtained classically from the eigenfunctions of the (dilaton-dressed) Klein-Gordon operator in the two-dimensional black hole background \cite{Dijkgraaf:1991ba} (in the large $k$ limit). 
They are parameterized by two numbers $\omega$ and $p$.\footnote{Here, we focus on the sector that descends from the continuous representations of $SL(2,\mathbb{R})$. The studies of the discrete states can be found in \cite{Distler:1991wr}\cite{Yogendran:2015upa}\cite{Yogendran:2018ikf}. See also \cite{Giveon:2019gfk}\cite{Giveon:2020xxh} in the context of thermal NS5-branes.} It has $\Delta = \bar{\Delta} = \frac{p^2-\omega^2+1}{4(k-2)}$. More explicitly, 
\begin{align}
U^p_{\omega} &= -\frac{\Gamma^2(\nu_+)}{\Gamma(1-i\omega)\Gamma(-ip)} e^{-i\omega t}(\sinh\rho)^{-i\omega} F(\nu_+,\nu_-^*;1-i\omega;-\sinh^2\rho) \cr
V^p_{\omega} &= -\frac{\Gamma^2(\nu_+^*)}{\Gamma(1+i\omega)\Gamma(+ip)} e^{-i\omega t}(\sinh\rho)^{i\omega} F(\nu_+^*,\nu_-;1+i\omega;-\sinh^2\rho)  \cr
\nu_{\pm} &= \frac{1}{2} + i\left(\frac{p\pm \omega}{2}\right)
\end{align}
and their linear combinations depending on the boundary conditions. Here the boundary conditions are imposed at the event horizon: $U_\omega^p$ vanishes at $u=0$ (as $\sim u^{-i\omega}$) and $V_{\omega}^p$ vanishes at $v=0$ (as $\sim (-v)^{i\omega}$).

Note that without specifying the boundary condition, we have twice as many vertex operators as in the Euclidean two-dimensional black hole. In the Euclidean black hole, in contrast, the regularity at the tip of the cigar relates the two modes. In addition, we have to impose the Virasoro constraint to prescribe the physical model. In two-dimensional string theory (i.e. without extra dimensions, or $k=\frac{9}{4}$), it gives the massless condition $\omega = \pm p$, which would have been called ``tachyon" in critical string theory (but offset $\frac{1}{4(k-2)}$ becomes exactly $1$ at $k=\frac{9}{4}$ so it is massless). 
The excitation in the internal space can make the dispersion massive.

The Euclidean two-dimensional black hole has been much studied. It is dual to the sine-Liouville theory (so-called Fateev-Zamolodchikov-Zamolodchikov duality \cite{FZZ}\cite{Hori:2001ax}\cite{Hikida:2008pe}\cite{Giveon:2015cma}), which can interpreted as a T-duality along the Euclidean time circle. Several correlation functions were computed in \cite{FZZ}, and a salient feature is that the winding number is not conserved.
In \cite{Dijkgraaf:1991ba}\cite{Giveon:1991sy}, it was suggested that the Lorentzian version of this duality swaps the region outside of the event horizon and the region inside the singularity.

\section{Probe membrane dynamics}
Let us now review the probe D0-brane dynamics in the two-dimensional black hole in string theory. The dynamics is effectively described by the Dirac-Born-Infeld action: 
\begin{align}
I &= \int ds e^{-\Phi} \cr
 & = \int dt \cosh \rho \sqrt{ \tanh^2 \rho - \dot{\rho}^2} \ .  
\end{align}
In the second line, we have used the world line diffeomorphism to impose the static gauge and the dot denotes the time derivative.

Let us introduce several different coordinates that make the effective action more suggestive. If we introduce the ``Tachyon coordinate" $\sinh \rho = e^{-T}$, then we obtain the effective action
\begin{align}
I & = \int dt e^{-T}\sqrt{1-\dot{T}^2} \ ,
\end{align}
which resembles the effective action for the open string tachyon condensation. This is what is known as the ``tachyon-radion correspondence" \cite{Kutasov:2004dj}. The corresponding boundary states in string theory was constructed in \cite{Nakayama:2004yx}\cite{Nakayama:2004ge}\cite{Nakayama:2005pk}. In analogy with the open string tachyon condensation, it suggests that the D0-brane radiates and disappears into nothing from the viewpoint of the distant (asymptotic) observer.

The other coordinate useful for the local observer is the Kruskal-like coordinate introduced in \eqref{Kruskal}. In this coordinate, the metric coupling is canceled against the dilaton coupling, resulting in the  Dirac-Born-Infeld action of the form
\begin{align}
I = \int dt \sqrt{ \dot{u} \dot{v}} \ ,
\end{align}
which is the same as the one in the flat two-dimensional Minkowski space-time. It is remarkable to note that the probe D0-brane does not see any singular behavior at the event horizon $uv=0$ or even at the singularity located at $uv=1$. 

The solution of the equations of motion of the probe brane in the Kruskal-like coordinate is given by a straight line $u=u_0, v=v_0$ \cite{Yogendran:2004dm}. In the original coordinate, it is given by
\begin{align}
\sinh \rho \cosh (t-t_0) = \sinh \rho_0 , 
\end{align}
where we have assumed that the trajectory is time-like. In the original coordinate suitable for the distant (asymptotic) observer, the D0-brane does not seem to cross the event horizon, but in the Kruskal coordinate, we see that it crosses the event horizon and even the singularity (without feeling any violence). 

The corresponding boundary states were proposed in \cite{Nakayama:2006qm}. The most relevant one would be the boundary state that is absorbed by the black hole. It is given by the boundary wavefunction:
\begin{align}
\Psi(\omega,p) = B(\nu_+,\nu_-) \Gamma\left(1+\frac{ip}{\tilde{k}}\right) e^{-i\omega t_0} \left(e^{-ip\rho_0} - \frac{\cosh(\pi\frac{p-\omega}{2})}{\cosh(\pi\frac{p+\omega}{2})} e^{ip\rho_0} \right) \ . 
\end{align}
Here $\tilde{k} = k-2$ for the bosonic case (or $\tilde{k} = k$ in the supersymmetric case).
The boundary state suggests that the radiation from the D0-brane (when $\tilde{k}>1$ or $k>3$) seems to carry away the entire energy seen from the asymptotic observer. The boundary state is constructed in terms of the analytic continuation of the Euclidean theory and it is appropriate to describe the motion of the D0-brane outside of the event horizon. It is not obvious, however, if it includes the information inside the horizon or the singularity.

The main goal of this paper is to find an analogue of the D0-brane dynamics of the two-dimensional black hole in the large $D$ black hole.
For this purpose, let us consider general membrane dynamics in the large $D$ black hole. Since we do not necessarily work in the string theory, the membrane we consider may not be a D-brane, but we assume that the effective dynamics is governed by the Nambu-Goto action (i.e. action is given by the world volume).

Suppose a $p$-brane wraps around $p$ cycle of the $(D-2)$ dimensional sphere, which behaves like a particle in the reduced two-dimensional space-time. The effective motion in the two-dimensional space-time is described by  
\begin{align}
I = \int dt e^{-\frac{2 \Phi p}{D-2}} \sqrt{\tanh^2 \rho - \dot{\rho}^2} \ . 
\end{align}
The factor of the dilaton is determined by the dimensionality of the wrapped cycles.\footnote{We assume that the motion inside the sphere is  (marginally) stabilized. It may require other degrees of freedom on the membrane (such as flux) or replacing the sphere with a manifold with non-trivial topology.}
If we take $\lim_{D \to \infty} \frac{p}{D-2} = \frac{1}{2}$, then we obtain the large $D$ analogue of the D0-brane action in the two-dimensional string black hole. We, therefore, expect that such a membrane does not see the horizon as well as the singularity in the probe dynamics.

Let us also study the effective dynamics of the wrapped membrane that becomes an effective string in the two-dimensional black hole. Suppose $(p+1)$-brane wraps around $p$ cycle of the  $(D-2)$ dimensional sphere, which behaves like a string in the reduced two-dimensional space-time

 The effective dynamics of the string in the two-dimensional space-time is described by the induced Nambu-Goto action
\begin{align}
I = \int dt d\sigma e^{-\frac{2 \Phi p}{D-2}} \sqrt{-\det \gamma} \ , 
\end{align}
where $\gamma$ is the induced metric on the world sheet (i.e. $\gamma_{ab} d\sigma^a d\sigma^b = -2\frac{dUdV}{1-UV}$). Depending on the choice of $p$, we have some options.

If we take $\lim_{D \to \infty} \frac{p}{D-2} = \frac{1}{2}$ as above, we obtain the effective D1-brane dynamics in the two-dimensional black hole background (with or without world volume flux). The corresponding D1-brane dynamics in the two-dimensional black hole was studied in \cite{Nakayama:2007sb}. Some D1-barne solutions could have been obtained by the formal T-duality in the time-like isometry, but as is well-known, the T-duality in the time-like direction gives an imaginary electric flux.

There is one more interesting limit, which has no analogy with the D1-brane. Suppose, alternatively, we take $\lim_{D \to \infty} \frac{p}{D-2} = 1$, then the effective string action becomes that of the Nambu-Goto action in the flat two-dimensional  Minkowski space-time
\begin{align}
I = \int dt d\sigma \sqrt{-\det \gamma^0} \ , 
\end{align}
where $\gamma^0_{ab} d\sigma^a d\sigma^b = -2 dUdV$ is the induced metric for the flat Minkowski space-time.

In this way, the effective string dynamics from the wrapped membrane (around the entire $D-2$ sphere) reduces to the dynamics of free string in the two-dimensional Minkowski space. In particular, it does not see the black hole singularity.
Apart from the collapsed string (i.e. particle limit), folded strings or extended strings that occupy the entire $uv$ plane may exist.\footnote{Fundamental strings in two-dimensional string background has been studied in the literature \cite{Maldacena:2005hi}\cite{Balthazar:2018qdv}.} 
 Note that our effective string dynamics is different from that of the fundamental string in the two-dimensional black hole background. Apparently, there is no analogue in two-dimensional string theory.

\section{Beyond the singularity}
As we have seen, a particular probe membrane, which corresponds to D0-brane in a two-dimensional black hole, does not see any singularity as far as we are concerned with the effective action. It may then require serious consideration about what happens after crossing the singularity. 

The singularity appears at $uv=1$ in the metric
\begin{align}
ds^2 &= -2k \frac{du dv}{1-uv} \ . 
\end{align}
There the curvature becomes infinity and it shows the geodesic incompleteness. We, however, imagine that the metric can be continued beyond $uv=1$  by the same formula. Beyond the singularity (i.e. $uv>1$), we may use the coordinate $u= \cosh\rho e^{-t}, v= \cosh\rho e^{t}$. In this coordinate, the metric reads
\begin{align}
ds^2 & = k (-\coth^2\rho dt^2 + d\rho^2)     
\end{align}
and it has a time-like singularity at $\rho=0$. The nature of the singularity is different from the black hole singularity although they are located at the same place at $uv=1$. It may be regarded as a negative mass singularity. Note that in this coordinate, the time goes ``sideways" (from left to right) and the interpretation of time and space is exchanged across the singularity.

The continuation of the dilaton requires extra care because the naive continuation of $e^{-2\Phi}  = (1-uv)$ leads to a complex value of $\Phi$ when $uv>1$ and does not admit any physically reasonable interpretation there. We, here, propose that it is rather continued to $e^{-2\Phi}  = |1-uv|$, which admits a physical interpretation with a real value of the dilaton field even when $uv>1$. In order to accept this proposal, we, first of all, note that it solves the same equations of motion because the constant shift of $\Phi$ is a moduli of the equations of motion and it can be shifted by an imaginary unit $\frac{\pi}{2} i$ to change the sign of $e^{-2\Phi}$. In \cite{Dijkgraaf:1991ba}\cite{Giveon:1991sy}, this shift seems implicitly assumed through the Euclidean continuation.

With this continuation, the dilaton beyond the singularity  in the $(t, \rho)$ coordinate is given by
\begin{align}
e^{-2\Phi} =+\sinh^2\rho \ .
\end{align}
It can be uplifted to the $D$-dimenional metric
\begin{align}
ds^2_D = k (-\coth^2 \rho dt^2 + d\rho^2) + r_0^2 (\sinh^2\rho)^{\frac{2}{D-2}} d\Omega_{D-2}^2  
\end{align}
in the large $D$ limit. If we had not shifted the imaginary part of the dilaton, the $D$-dimensional uplift would have been less clear.

Now let us discuss the fate of the wrapped probe membrane (D0-brane) after crossing the singularity. We have already noted that the probe membrane moves in the $(u,v)$ plane as if the space-time is flat due to the cancellation between the metric coupling and the dilaton coupling. This is the same at the moment of crossing the singularity or beyond the singularity.

However, one thing to be noted is that beyond the singularity, space and time flip. This is more drastic than when we cross the event horizon, where the Killing ``time" becomes space-like. Here, the time-like trajectory is continued to be a space-like trajectory. Therefore, the continuation of the trajectory suggests that the wrapped probe membrane becomes an S-brane (space-like instantaneous object \cite{Gutperle:2002ai}) that stems from the negative mass singularity. In the $(t,\rho)$ coordinate, the trajectory is
\begin{align}
\cosh\rho \sinh(t-t_0) = \cosh\rho_0 \ . 
\end{align}
Due to the flip of space and time, it is not immediately obvious how the observer actually perceives when he or she crosses the singularity.\footnote{Some readers might recall what happens after crossing the singularity in the movie ``Interstellar".} 

If the trajectory is null, the continued trajectory remains null, so the physically sensible continuation may be possible. Of course, if an object has a time-like trajectory at some point, it requires infinite energy to make the trajectory null, so it may not be a practically good idea to accelerate himself or herself inside the black hole to avoid the situation described in the previous paragraph.

In the case of the effective string dynamics, it reduces to that of the D1-brane in the two-dimensional black hole (that can see the singularity) or the two-dimensional string in the flat Minkowski space-time. The latter does not see any singularity of the black hole. Beyond the singularity, the time direction and the spatial direction swap, but since the string is extended, the more elaborate continuation of the notion of time and space would be possible. Of course, since the human being is a local object, we do not know precisely how the extended object perceives space and time, and, hence, not much can be said here.

\section{Discussions}
In this paper, we have shown that the black hole singularity can be circumvented by a particular probe observer in the large $D$ limit. It should be exciting to see if the mechanism can survive with the $1/D$ correction and beyond the probe approximation.  One may even try $D=4$ and $p=1$ to test the idea in the observed black holes in our universe. In the context of the string theory, this is more than wishful thinking because, in the Euclidean theory, the singularity and the horizon are dual each other, and both make sense. It is equally likely that the would-be singularity may not be singular at all in string theory.

While it is speculative if the same mechanism can be employed in general relativity at finite $D$, one lesson here is that in the warped compactification, the probe can feel various metrics depending on the dimensionality. This is a generic feature in higher dimensional gravity as well as in string theory and it is worth pursuing further.

In string theory, there are more interesting situations where the probe branes witness enhanced symmetries due to the cancellation between the metric and the dilaton. The most notable example is the probe D$p$-brane in the D$k$-brane background with $p+k=6$. In this circumstance, the probe D$p$-brane shows the enhanced conformal symmetry (i.e. ``AdS isometry") even though the bulk geometry does not show such a symmetry.\footnote{The details of such mechanism with more interesting examples can be seen in unpublished work \cite{NR}.}

At the technical level, it is important to construct the exact boundary states that correspond to the space-like brane or null brane that is obtained after crossing the singularity. The null brane can be used to probe the chaotic behavior inside the black hole \cite{Chen:2021emg}\cite{Shenker:2013pqa}. It is also of interest to study the large $D$ limit of other black hole solutions or cosmological solutions in relation to the two-dimensional string theory and D-branes there.

\section*{Acknowledgements}
 This work is in part supported by JSPS KAKENHI Grant Number 21K03581. The author would like to thank Soo-Jong Rey for the correspondence and the discussions on \cite{NR} at KEK.

\end{document}